\documentclass[prx,aps,twocolumn,preprintnumbers,amsmath,amssymb,superscriptaddress,showpacs,longbibliography]{revtex4-1}
\usepackage[colorlinks,bookmarks=true,citecolor=blue,linkcolor=red,urlcolor=blue,citecolor=blue]{hyperref}

\usepackage{graphicx}
\usepackage{subfigure}
\usepackage{amssymb,amsmath}
\usepackage{color}
\usepackage{enumitem,kantlipsum}
\usepackage{ulem}
\usepackage{balance} 
\usepackage{lineno}
\newcommand{\rzz}{R_{\mathrm ZZ}}
\newcommand{\rx}{R_{\mathrm X}(\theta_h)}

\begin{document}
\switchlinenumbers
\title{Simulation of IBM's kicked Ising experiment with Projected Entangled Pair Operator}

\author{Hai-Jun Liao}
\email{navyphysics@iphy.ac.cn}
\affiliation{Beijing National Laboratory for Condensed Matter Physics and Institute of Physics, Chinese Academy of Sciences, Beijing 100190, China.}
\affiliation{Songshan Lake Materials Laboratory, Dongguan, Guangdong 523808, China.}

\author{Kang Wang}
\affiliation{Beijing National Laboratory for Condensed Matter Physics and Institute of Physics, Chinese Academy of Sciences, Beijing 100190, China.}
\affiliation{School of Physical Sciences, University of Chinese Academy of Sciences, Beijing 100049, China.}

\author{Zong-Sheng Zhou}
\affiliation{Beijing National Laboratory for Condensed Matter Physics and Institute of Physics, Chinese Academy of Sciences, Beijing 100190, China.}

\author{Pan Zhang}
\email{panzhang@itp.ac.cn}
\affiliation{CAS Key Laboratory for Theoretical Physics, Institute of Theoretical Physics, Chinese Academy of Sciences, Beijing 100190, China}
\affiliation{School of Fundamental Physics and Mathematical Sciences, Hangzhou Institute for Advanced Study, UCAS, Hangzhou 310024, China}
\affiliation{International Centre for Theoretical Physics Asia-Pacific, Beijing/Hangzhou, China}

\author{Tao Xiang}
\email{txiang@iphy.ac.cn}
\affiliation{Beijing National Laboratory for Condensed Matter Physics and Institute of Physics, Chinese Academy of Sciences, Beijing 100190, China.}
\affiliation{Beijing Academy of Quantum Information Sciences, Beijing, China.}
\affiliation{School of Physical Sciences, University of Chinese Academy of Sciences, Beijing 100049, China.}

\begin{abstract}
We perform classical simulations of the 127-qubit kicked Ising model, which was recently emulated using a quantum circuit with error mitigation [Nature 618, 500 (2023)]. Our approach is based on the projected entangled pair operator (PEPO) in the Heisenberg picture. Its main feature is the ability to automatically identify the underlying low-rank and low-entanglement structures in the quantum circuit involving Clifford and near-Clifford gates.

We assess our approach using the quantum circuit with 5+1 trotter steps which was previously considered beyond classical verification. We develop a Clifford expansion theory to compute exact expectation values and use them to evaluate algorithms. The results indicate that PEPO significantly outperforms existing methods, including the tensor network with belief propagation, the matrix product operator, and the Clifford perturbation theory, in both efficiency and accuracy. In particular, PEPO with bond dimension $\chi=2$ already gives similar accuracy to the CPT with $K=10$ and MPO with bond dimension $\chi=1024$. And PEPO with $\chi=184$ provides exact results in $3$ seconds using a single CPU.

Furthermore, we apply our method to the circuit with 20 Trotter steps. We observe the monotonic and consistent convergence of the results with $\chi$, allowing us to estimate the outcome with $\chi\to\infty$ through extrapolations. We then compare the extrapolated results to those achieved in quantum hardware and with existing tensor network methods. Additionally, we discuss the potential usefulness of our approach in simulating quantum circuits, especially in scenarios involving near-Clifford circuits and quantum approximate optimization algorithms. Our approach is the first use of PEPO in solving the time evolution problem, and our results suggest it could be a powerful tool for exploring the dynamical properties of quantum many-body systems.
\end{abstract}

\maketitle

\section{Introduction}
A recent experiment~\cite{Nature} provided evidence supporting the utility of quantum computing before fault tolerance. This was accomplished through the zero-noise extrapolated quantum simulation of the kicked Ising model, using up to 127 qubits. By comparing with the matrix product state (MPS) and isometric tensor network state (isoTNS) simulations~\cite{Nature}, it was shown that IBM's quantum hardware delivered more accurate results when the expectation can be verified using exact values with $5$ trotter steps. 
 
Recently, several novel classical algorithms have emerged, aiming to challenge the efficacy of quantum simulations. These include the belief propagation tensor network state (BP-TNS)~\cite{tindall2023efficient}, the Heisenberg matrix product operator (MPO)~\cite{Zaletel23}, the Clifford perturbation theory (CPT)~\cite{Chan23}, the 31-qubit subset simulation~\cite{kechedzhi2023effective}, and observable’s back-propagation on Pauli paths (OBPPP)~\cite{shao2023simulating}. These classical algorithms can compute more accurately the expectation values in the verifiable regime with 5 Trotter steps utilizing only moderate computational resources. However, the results of various methods exhibited around a $20\%$ deviation for a quantum circuit with $20$ Trotter steps within the regime of $\pi/8 \le \theta_h \le 3\pi/8$~\cite{Zaletel23}. This discrepancy suggests that the accurate results with $\theta_h$ away from $\pi/2$ remain unclear, and it is difficult to access the accuracy of IBM's quantum hardware in that parameter regime.

 In this work, we map the expectation computation to the contraction problem of a tensor network with the observable operator in the middle of the network. We propose to contract the tensor network based on the PEPO representation of the Heisenberg evolution operator. It applies the single-qubit and two-qubit rotation gates to the operator layer by layer from the middle to the boundary of the tensor network. Compared with other tensor-network methods, our approach can automatically detect the light-cone structure (i.e., the funnel shape) of the tensor network, the intrinsic low-rank structures circuit involving Clifford ZZ-rotation gates, and the low-entanglement structure when the X-rotation gates are close to the Clifford limit. It completely avoids the use of long-range operators and swap operations. Consequently, our approach can accurately simulate IBM's 127-qubit quantum circuit.

 To quantitatively demonstrate the performance of our method, we use IBM's kicked Ising model with $5+1$ Trotter steps (corresponding to Fig. 4a in Ref~\cite{Nature}) as a benchmark. For this particular system, we propose an exact Clifford expansion theory to simplify the quantum circuit and manage to obtain exact results for different expectation values. Remarkably, the quantum circuit with $5+1$ Trotter steps has been considered not in the classically verifiable regime and thus has not been used for evaluating algorithms in previous works due to the lack of exact results. Based on this benchmark, we further show that our method is significantly more accurate than the quantum hardware with error mitigations and other existing tensor network algorithms. Our approach reaches a rounding error in less than $3$ seconds on a single CPU.

 The paper is organized as follows. In Sec.~\ref{sec:model}, we describe the kicked Ising model and IBM's quantum circuits. In Sec.~\ref{sec:method}, we introduce our PEPO method and compare it with other tensor network approaches. In Sec.~\ref{sec:6}, we apply the method to IBM's quantum circuit and present the results for the quantum circuits with $5+1$ and 20 Trotter steps. We conclude in Sec.~\ref{sec:con}.

\section{IBM's kicked Ising experiment} \label{sec:model}

\begin{figure}[tp]
\centering
\includegraphics[width=0.45\textwidth]{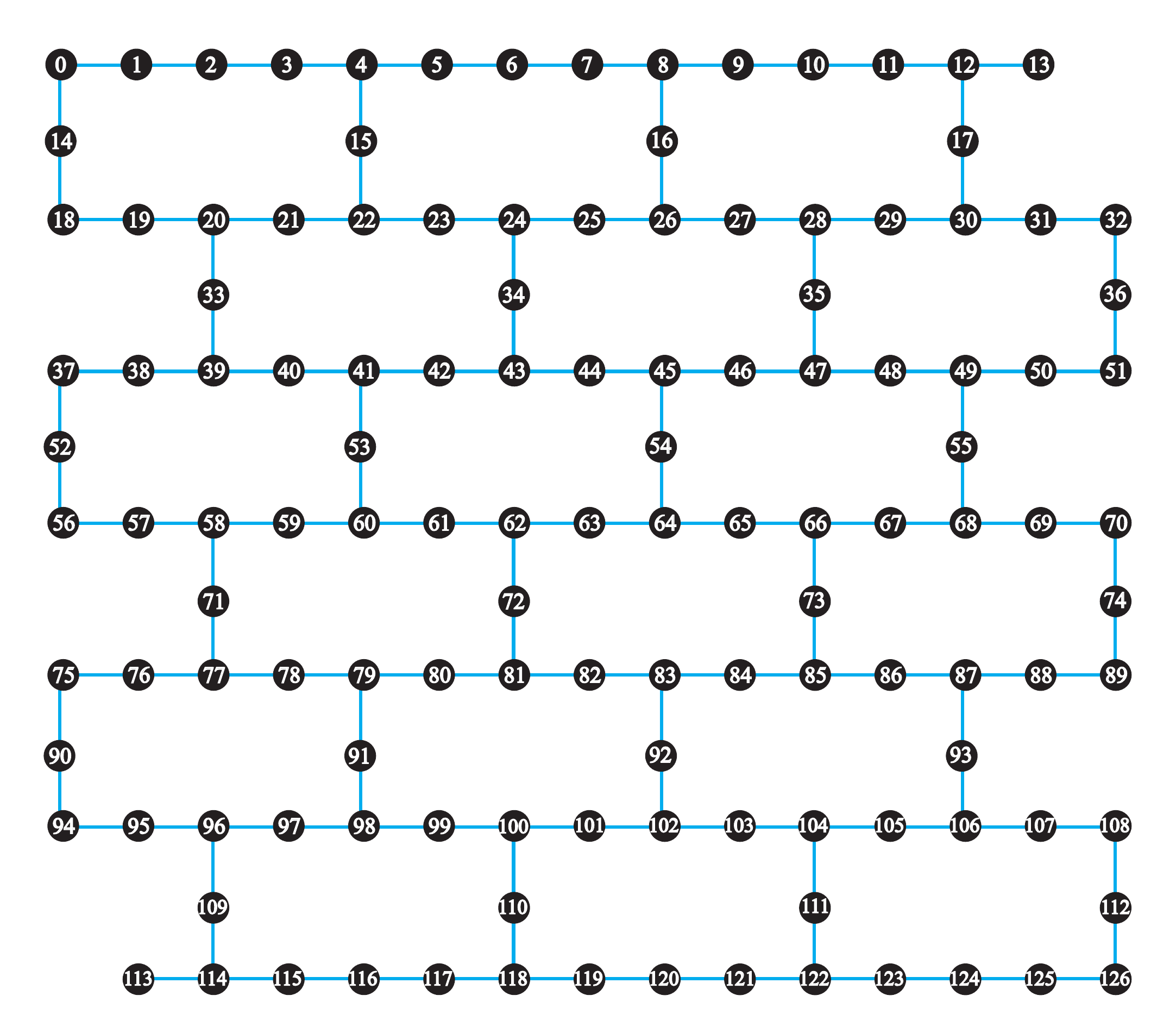}
\caption{Layout of IBM's 127-qubit quantum processor.}
 \label{fig:lattice}
\end{figure}
A recent experiment was carried out by IBM in simulating the dynamics of the transverse-field Ising model (kicked Ising model) on a two-dimensional heavy-hexagon lattice (as illustrated in Fig.~\ref{fig:lattice}) using a 127-qubit quantum circuit~\cite{Nature}. The experiment demonstrated evidence for the utility of quantum computing before fault tolerance using error mitigation. The quantum circuit simulates the kicked Ising model with $T$ steps of unitary evolutions 
\begin{equation}
    U_T(\theta_h) = \left[\rzz\rx\right]^{T},
\end{equation}
where in each step the unitary evolution is composed of the Clifford gates on each edge $\langle i,j \rangle$, and the X-rotation gates on each qubit
\begin{eqnarray}
    \rzz &= \prod_{\langle i,j \rangle}\exp \left({\rm i} \frac{\pi}{4}Z_{i}Z_{j} \right), \\
    \rx&=\prod_{i}\exp \left(-{\rm i} \frac{\theta_{h}}{2}X_{i}\right).
\end{eqnarray}
Notice that the X-rotation gates are not Clifford except at $\theta_h = k\pi/2$ with $k$ an integer.

In Ref.~\cite{Nature}, the authors simulated the expectation values using the quantum hardware with error mitigation and compared the results against tensor network algorithms on \textit{three settings}:\\
\begin{enumerate}[labelindent=0pt,leftmargin=12px]
\item The circuit is shallow, with depth restricted to $T=5$ Trotter steps, and the observable is carefully chosen such that the expectation value can be computed exactly. By comparison with the exact results, Ref.~\cite{Nature} shows that the results of quantum hardware are very close to the exact ones, much more accurate than results obtained using MPS and isoTNS even with large bond dimensions.

\item The circuit has $5$ Trotter steps with an additional layer of $R_X$ gates and effectively simulates the time evolution after $6$ Trotter steps. So we term it as a circuit with $5+1$ Trotter steps. In this case, the expectation values are much more difficult to compute than in the system with $5$ steps, and the previous studies~\cite{Nature, tindall2023efficient, Zaletel23, Chan23} consider the circuit beyond exact verification.

\item The circuit is deep, with $T=20$ Trotter steps. This setting is not classically verifiable. In Ref.~\cite{Nature}, a large deviation between the experimental data and the result of MPS and isoTNS is observed.
\end{enumerate}

 It was reported that in settings 2 and 3, the hardware results of expectation values significantly deviate from the tensor network results, demonstrating the utility of near-term quantum devices using error mitigations in the regime of strong entanglements when canonical tensor network methods break down. Soon after Ref.~\cite{Nature} was published, several novel classical algorithms have been proposed~\cite{tindall2023efficient, Chan23, Zaletel23}, reporting that the advanced tensor network algorithms outperform the canonical tensor network methods used in Ref.~\cite{Nature} in setting 1. However, it has been noted that there is a large discrepancy among different methods in setting 3. The accuracy of the hardware results in settings 2 and 3 also remains unknown.

\section{Heisenberg PEPO evolution\label{sec:method}}
The time-dependent expectation of an operator $\langle \hat{O}(t)$ can be calculated in the Schr\"odinger picture
\begin{equation} 
\langle \hat{O}(t) \rangle  = \langle \Psi(t)| \hat{O} | \Psi(t)\rangle,
\end{equation} 
or in the Heisenberg picture
\begin{equation} 
\langle \hat{O}(t) \rangle  = \langle \Psi | \hat{O}(t) | \Psi \rangle.
\end{equation} 
Where $| \Psi(t)\rangle =  e^{-i\hat{H}t} | \Psi \rangle$ is the time-dependent quantum state, and $\hat{O}(t)= e^{+i\hat{H}t} \hat{O}e^{-i\hat{H}t}$ is the time-dependent Heisenberg operator. 
In both pictures, the quantum state $| \Psi(t)\rangle$ or the time-dependent Heisenberg operator $\hat{O}(t)$) can be represented using a tensor network such as MPS or MPO, and are evolved using an algorithm such as the time-evolving block decimation (TEBD)~\cite{TEBD1,TEBD2}, simple-update~\cite{simple-update} and full-update~\cite{full-update} methods. When the entanglement of the tensor network is large enough, one needs to adopt approximate truncations on the virtual bond of the tensor network to reduce the computational complexity of the algorithm. In the case of IBM's kicked Ising experiments, MPS~\cite{Nature}, isoTNS~\cite{Nature}, and BP-TNS~\cite{tindall2023efficient} methods belong to the Schrodinger picture, and the MPO method of~\cite{Zaletel23} is conducted in the Heisenberg picture. 

Both pictures can be regarded as different contraction schemes of a (d+1) tensor network corresponding to the time evolution of the d-dimensional quantum system. In the case of IBM's kicked Ising model, the qubits locate on a two-dimensional heavy-hexagon lattice, then the corresponding tensor network of computing expectation of an observable operator is a three-dimensional tensor network $\mathcal T$ with the observable in the middle of the tensor network. The expectation value is computed by contracting the three-dimensional tensor network. 

In the Schr\"odinger picture, the contraction is carried out from the boundary (corresponding to the initial state) to the middle (corresponding to the observable); while in the Heisenberg picture, the contraction is carried out from the middle to the two boundaries, which means at each time, the time evolution operator and its conjugate are applied simultaneously. The two pictures are mathematically equivalent if no approximation is introduced. 

However, in practice, each picture has its advantages and disadvantages. The tensor network state in the Schr\"odinger picture typically has a much lower space complexity than the tensor network operator in the Heisenberg picture, allowing it to employ a much larger virtual bond dimension and obtain more accurate results. The Heisenberg picture, on the other hand, exploits the intrinsic structure in the form of $UOU^\dagger$ and may greatly simplify the tensor network calculation in some situations, e.g. when the entanglements generated by the unitary and its inverse transformations partially cancel each other. 

In this work, we propose to represent the observable of the three-dimensional tensor network $\mathcal T$ using PEPO in the Heisenberg picture and contract $\mathcal T$ by evolving PEPO from the middle to the two boundaries. The compression of the tensors is performed using the simple-update~\cite{simple-update}, together with an exact contraction of the final tensor network to obtain the expectation value. Compared with the BP-TNS approach~\cite{tindall2023efficient}, which introduces uncontrolled approximations due to the message passing, the error of our method is controlled via the error of singular value decompositions and can reproduce the exact results at $\chi \rightarrow \infty$.

\begin{figure*}[tp]
\centering
\includegraphics[width=0.45\textwidth]{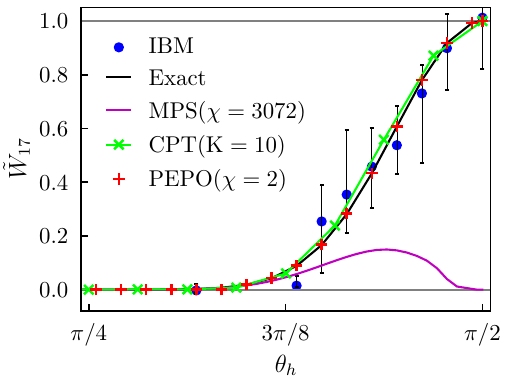}
\includegraphics[width=0.45\textwidth]{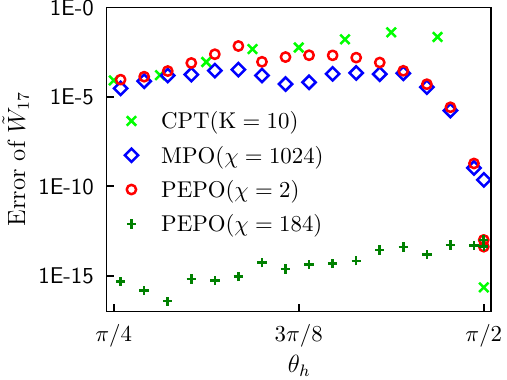}
\caption{\textit{Left}: The expectation value of the modified weight-17 stabilizer $\tilde{W}_{17}$ obtained by PEPO, MPS~\cite{Nature}, CPT~\cite{Chan23}, the IBM's quantum hardware with error mitigation (IBM)~\cite{Nature}, and compared against exact results, on the quantum circuit with 5+1 Trotter steps ($5$ Trotter steps with an addition rotation gates, corresponding to Fig.4(a) in Ref.~\cite{Nature}). 
\textit{Right}: The absolute errors with respect to the exact results.
The computation time of the PEPO method with $\chi=184$ in obtaining a data point is less than $3$ seconds using a single Intel Xeon Gold 6326 CPU. 
}
\label{fig:shallow}
\end{figure*}

The computational cost at each step of evolution is $\mathcal{O}(L\chi^4)$, with $L=144$ the number of edges of the heavy-hexagon lattice and $\chi$ the virtual bond dimension. The computational cost of the exact tensor network contraction at the final step is $\mathcal{O}(\chi^6)$. At first glance, the computational complexity with respect to the bond dimension $\chi$ looks much higher than MPS~\cite{Nature}, isoTNS~\cite{Nature}, BP-TNS~\cite{tindall2023efficient} and MPO~\cite{Zaletel23}, we observe that the computation is more effective than MPS, isoTNS, BP-TNS and MPO for two reasons. 

\begin{enumerate}[wide, labelindent=0pt,leftmargin=11px]

\item  PEPO reflects the two-dimensional geometry of the heavy-hexagon lattice, so all the time-evolution operators are local. In contrast, in the one-dimensional MPO representation~\cite{Zaletel23}, some of the time evolution operators are long-ranged, so the use of  SWAP operations is inevitable in MPO, reducing its efficiency and accuracy. 

\item In addition to the cancellation effect of conjugate unitary gates, PEPO in the Heisenberg picture can automatically catch the intrinsic low-rank structure due to the presence of Clifford $ZZ$ rotations gates and the approximate low-entanglement structures induced by the $X$ rotation gates with $\theta_h$ close to $\pi/2$. This can dramatically reduce the computation cost and enhance the algorithm's effectiveness. As a simple example, PEPO with $\chi=1$ can obtain exact results at the Clifford points, for instance, at $\theta_h=\pi/2$, as illustrated in Fig.~\ref{fig:shallow}(b). In contrast, MPS, isoTNS and BP-TNS methods have no chance to meet the form of $UOU^\dagger$ hence can not detect the low-rank structure at all, giving completely wrong results at the Clifford points~\cite{Nature}.

\end{enumerate}

\section{Results} \label{sec:6}

\subsection{Circuit with $5+1$ Trotter steps} 

We first present the results obtained on shallow circuits, focusing on Setting 2: the shallow circuits with $5+1$ Trotter steps. We choose Setting 2 because it is more difficult to compute than Setting 1, so comparing different algorithms can be demonstrated more clearly. In the previous studies, only setting 1 was used to compare errors because setting 2 was considered unverifiable. 

Here we show that setting 2 is also verifiable. We propose an exact Clifford expansion theory (CET) to reduce the depth of the circuit, followed by an exact contraction of the corresponding tensor network using the tensor slicing technique~\cite{3DIsing,Pan}. Details about CET can be found in the Appendix. This technique allows us to rigorously compute the expectation value of $\tilde{W}_{17}$ for this particular circuit with 5+1 steps. We do not invoke this technique in the PEPO tensor-network calculations.

 Figure~\ref{fig:shallow} shows the calculated results. On the left panel, we see that the IBM measurement, MPS, and CPT results deviate clearly from the exact results, while our PEPO results obtained just with a small bond dimension $\chi=2$ already agree better with the exact values. The right panel of Fig.~\ref{fig:shallow} compares the absolute errors of the results obtained with different algorithms, showing clearly that PEPO with $\chi=2$ has similar accuracy as MPO with $\chi=1024$ and CPT with $K=10$. It indicates that our Heisenberg PEPO method can automatically detect the intrinsic structure of Clifford gates. Furthermore, by taking $\chi=184$, we find that the errors of the PEPO results already fall below the rounding error of the double-precision floating numbers. Precisely at the Clifford point with $\theta_h=\pi/2$, the error of the $\chi=2$ PEPO result drops to zero, indicating that our approach perfectly catches the low-rank structure of the Clifford gates. The computation time for each point is less than $3$ seconds for $\chi=184$ using one CPU. In our PEPO calculation, we directly evolve the tensor-network operator using the simple update starting from the original circuit without using any information obtained from the manual Clifford expansions. It clearly shows that the PEPO method can detect the low-entanglement structure of the circuit automatically.

\subsection{Circuit with $20$ Trotter steps}\label{sec:20}

\begin{figure*}[tp]
\centering
\includegraphics[width=0.45\textwidth]{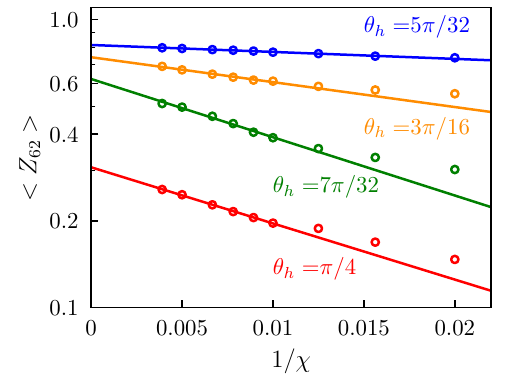}
\includegraphics[width=0.45\textwidth]{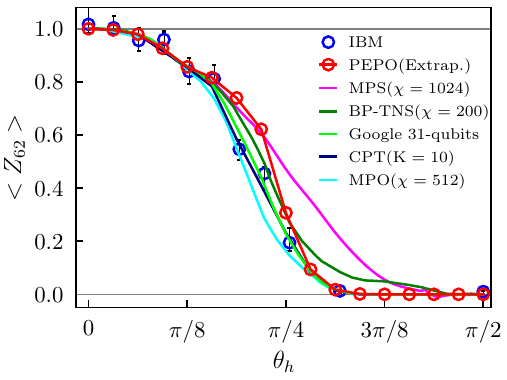}
\caption{\textit{Left}: Extrapolation of the expectation values of $\langle Z_{62}\rangle$ obtained by our method as a function of the inverse bond dimension $\chi$ on IBM's circuit with $20$ Trotter steps. The solid lines are extrapolated results using a function $b\,e^{-a/\chi}$ with two fitting parameters $a$ and $b$. The largest bond dimension we have calculated is $\chi=256$, which takes about $7$ hours using a single Intel Xeon Gold 6326 CPU to obtain a single data point.
\textit{Right}: The extrapolated value of $\langle Z_{62}\rangle$, PEPO (Extrap.), compared with the results of IBM's quantum hardware with error mitigation (IBM) and other numerical algorithms.
}
\label{fig:deep}
\end{figure*}

Here we conduct numerical experiments on deep circuits with $20$ Trotter steps. Figure~\ref{fig:deep} (left) shows the expectation value of $\langle Z_{62}\rangle$ computed using PEPO with different $\chi$. We find that $\langle Z_{62}\rangle$ converges very quickly with increasing $\chi$ and becomes nearly $\chi$  independent in the regimes $\theta_h \le \pi/8$ and $\theta_h \ge 5\pi/16$. In the intermediate regime, $\pi /8 <\theta_h < 5 \pi /16 $, $\langle Z_{62}\rangle$ shows visible variance with $\chi$ due to the rapidly increasing entanglement of PEPO with the Trotter steps. However, $\langle Z_{62}\rangle$ varies monotonically with increasing $\chi$, unlike the results obtained with MPO, in this regime, allowing us to reliably estimate the values of  $\langle Z_{62}\rangle$ by extrapolation to the limit $\chi\to\infty$. 

Figure~\ref{fig:deep} (right) compares our results with the IBM measurement data after error mitigation and those published by other calculations. In the regime $\theta_h\leq \pi/8$, the results of all approximate algorithms agree well with each other. In the regime $\theta_h > 5\pi/16$, the results of PEPO, Google 31-qubits, CPT, and MPO all converge to $0$, while isoTNS, BP-TNS and MPS results deviate from zero significantly. In this regime, the computation is pretty easy because the X-rotation gates are close to the Clifford limit. The deviation of the MPS, isoTNS and BP-TNS is because these methods can not detect the entanglement structures even in the near-Clifford limit. In the intermediate regime, $\pi/8<\theta_h<5\pi/16$, a discrepancy is observed between the results obtained with different methods. The classical simulation becomes challenging in this regime because the X-rotation gates deviate significantly from the Clifford limit, and the entanglement becomes strong. Notably, PEPO gives considerably greater values than other results in this regime. Before extrapolation, the PEPO results increase with increasing $\chi$. Their differences with the CPT  and IBM's measurement results also grow with increasing $\chi$. However, due to the strong entanglement and non-verifiable nature, we cannot tell which method is more accurate in this regime.

\section{Discussion and Conclusion} \label{sec:con}

We have developed an accurate and efficient approach for simulating the discretized dynamics of the kicked Ising model first investigated on a 127-qubit quantum circuit in Ref.~\cite{Nature}. Our algorithm is based on PEPO representation of the evolution operator in the Heisenberg picture. It automatically identifies the low-rank structure and low-entanglement structures in the circuit, which reduces the computational cost but increases the accuracy of the simulation. For the quantum circuit with 5+1 Trotter steps, which is previously considered not verifiable, we propose an exact Clifford expansion scheme to evaluate the expectation values exactly. This expansion theory outperforms all other simulation methods in this system. Furthermore, we find that the PEPO method with a bond dimension $\chi=2$ can already give similarly accurate results as CPT with $K=10$ and MPO with bond dimension $\chi=1024$. Finally, we apply the PEPO method to the deep circuit with $20$ Trotter steps. 

Our findings reveal the remarkable effectiveness of the Heisenberg PEPO method in the calculation of dynamical evolutions. This method shows promise for computing quantum system expectations, which is especially useful for applications like the QAOA system, which bears similarities to the kicked Ising model investigated in this study. We intend to delve further into this direction in the future.

\begin{acknowledgements}
An implementation of our algorithm can be found at~\cite{github}.
We thank Garnet Kin-Lic Chan and Tomislav Begu\u{s}i\'c for offering data in~\cite{CPT}, and thank Sajant Anand, Abhinav Kandala, and Michael Zaletel for offering data in~\cite{Zaletel23}.
This work is supported by the National Key Research and Development Project of China (Grants No.~2022YFA1403900 and No.~2017YFA0302901), 
the National Natural Science Foundation of China (Grants Nos.~11888101,~11874095, and~11974396), 
the Youth Innovation Promotion Association CAS (Grants No.~2021004), and the Strategic 
Priority Research Program of Chinese Academy of Sciences (Grant Nos.~XDB33010100 and~XDB33020300). 
\end{acknowledgements}
\bibliography{IBM127}

\onecolumngrid

\section*{Appendix: Abbreviations}
\begin{enumerate}
    \item BP-TNS: belief propagation tensor network state 
    \item CET: Clifford expansion theory
    \item CPT: Clifford perturbation theory
    \item isoTNS: isometric tensor network state 
    \item MPO: matrix product operator 
    \item MPS: matrix product state 
    \item OBPPP: observable’s back-propagation on Pauli paths
    \item PEPO: projected entangled pair operator
    \item TEBD: time-evolving block decimation
    \item QAOA: quantum approximate optimization algorithm
\end{enumerate}

\section*{Appendix: Clifford Expansions Theory for circuit simplifications}

In this section, we describe how to manually utilize the commutation relations and the structure of the Clifford quantum gates $R_{ZZ}$,non-Clifford gates $R_{X}(\theta_h)$, and Pauli operators to simplify the circuit and reduce the circuit depth. For the shallow circuit, this may enable the exact computation of observables after the simplification. Here we give some examples. We can reduce the computation of the expectation value of the Weight-10 ($W_{10}$) stabilizer after 5 Trotter steps as follows

\begin{eqnarray}
\langle W_{10} \rangle_{5} &=& \langle 0 | U^{\dagger}_{5}[\theta_h] (X_{13,29,31}Y_{9,30}Z_{8,12,17,28,32}) U_{5}[\theta_h] |0\rangle \nonumber\\
&=& \langle 0 | U^{\dagger}_{3}[\theta_h] R^{\dagger}_X \Big[ Z_{17}( c_h X_{13}Z_{12}+ s_hZ_{13}) (c_h^2 Y_{9}Z_{8} + c_h s_h Z_{9}Z_{10} -s^2_h Y_{10}Z_{11} -c_h s_h X_{9}X_{10}Z_{8}Z_{11}) \nonumber \\
&& (c^2_h X_{29}X_{31}Y_{30}Z_{28}Z_{32}  - c_h s_h X_{29}X_{30}Z_{28} -c_h s_h X_{30}X_{31}Z_{32} -s^2_h Y_{30})\Big]
 R_X \, U_{3}[\theta_h] |0\rangle,   
\end{eqnarray}

where $c_h=\cos \theta_h $ and $s_h =\sin \theta_h $.

The expectation of the Weight-17 ($W_{17}$) stabilizer after 5 Trotter steps can be computed as
\begin{eqnarray}
\langle W_{17} \rangle_{5} &=& \langle 0 | U^{\dagger}_{5}[\theta_h] (X_{37,41,52,56,57,58,62,79}Y_{75}Z_{38,40,42,63,72,80,90,91}) U_{5}[\theta_h] |0\rangle  \nonumber \\
&=& \langle 0 | U^{\dagger}_{4}[\theta_h] R^{\dagger}_X (-X_{52,56,57,58}Y_{37,41,62,75,79}Z_{53,59,61,71,76,78}) R_X \, U_{4}[\theta_h] |0\rangle  \nonumber \\
&=& \langle 0 | U^{\dagger}_{3}[\theta_h] R^{\dagger}_X \Big\{ 
X_{56}X_{57}(c_h  X_{37,52}Z_{38} +  s_h  Y_{52}) \nonumber \\
&& ( s^2_h  Y_{61}Z_{60} + c_h  s_h  X_{61}Y_{62} Z_{60,63,72} - c_h  s_h  Z_{61,62} +  c^2_h  X_{62}Z_{63,72} )\nonumber \\
&& ( s^2_h  Y_{76}Z_{77} + c_h  s_h  X_{75,76}Z_{77,90}  - c_h  s_h  Z_{61,62} +  c^2_h  X_{62}Z_{63,72} )\Big\} 
 \nonumber \\
&& \Big\{ \Big[ ( c^3_h  X_{41}Y_{58}Z_{40,42} -  c^2_h   s_h  X_{41,58,59}Z_{40,42,60} -  c^2_h   s_h  Y_{58}Z_{41,53} +  c^2_h   s_h  X_{53}Y_{41,58}Z_{40,42,60} \nonumber \\ 
&& -\ s^3_h  X_{58,59}Y_{53} + c_h   s^2_h  X_{58,59}Z_{41,53,60} + c_h   s^2_h  Y_{53,58}Z_{60} - c_h   s^2_h  X_{53,58,59}Y_{41}Z_{40,42} ) \nonumber\\
&& ( c_h   s^2_h  X_{71}Z_{77,78,79} - c_h   s^2_h  X_{71,78}Y_{79}Z_{80,91} -  c^2_h   s_h  X_{71,79}Z_{77,80,91} - s^3_h  X_{71}Y_{78}) \Big] \nonumber\\
&& + \ \Big[ (- c^3_h  X_{41,58}Z_{40,42} +  c^2_h   s_h  X_{58}Z_{41,53} -  c^2_h   s_h  X_{53,58}Y_{41}Z_{40,42,60} -  c^2_h   s_h  X_{41,59}Y_{58}Z_{40,42,60} \nonumber \\ 
&& - \ s^3_h  X_{59}Y_{53,58} + c_h   s^2_h  X_{59}Y_{58}Z_{41,53,60} - c_h   s^2_h  X_{58}Y_{53}Z_{60} - c_h   s^2_h  X_{53,59}Y_{41,58}Z_{40,42} ) \nonumber\\
&& ( c_h   s^2_h  Y_{78}Z_{77} -  c^2_h   s_h  Z_{78,79} +  c^2_h   s_h  X_{78}Y_{79}Z_{77,80,91} + c^3_h  X_{79}Z_{80,91}) \Big] \Big\} R_X \, U_{3}[\theta_h] |0\rangle.
\end{eqnarray}

The expectation of the modified Weight-17 ($\tilde{W}_{17}$) stabilizer can be reduced as follows
\begin{eqnarray}
\langle \tilde{W}_{17} \rangle_{5} &=& \langle 0 | U^{\dagger}_{5}[\theta_h] R^{\dagger}_X (X_{37,41,52,56,57,58,62,79}Y_{38,40,42,63,72,80,90,91}Z_{75}) R_X \, U_{5}[\theta_h] |0\rangle  \nonumber \\
&=& \langle 0 | U^{\dagger}_{4}[\theta_h] R^{\dagger}_X \Big\{ 
(-X_{52,56,57,58}Z_{53,59,61,71}( s^3_h  Y_{37,41} + c_h  s^2_h  X_{37,38}Y_{41} Z_{39} + c_h  s^2_h  X_{40,41}Y_{37}Z_{39}  \nonumber\\
&& + \ c_h  s^2_h  X_{41,42}Y_{37}Z_{43} +  c^2_h  s_h  X_{37,38,40,41} +  c^2_h  s_h  X_{37,38,41,42}Z_{39,43}  -  c^2_h  s_h  X_{40,42}Y_{37,41}Z_{39,43} \nonumber\\
&& - \ c^3_h  X_{37,38,40,42}Y_{41}Z_{43} ) ( s^2_h  Y_{75}Z_{76}  - c_h  s_h  Z_{75,90} + c_h  s_h X_{75,90}Z_{76,94} - c^2_h  Y_{90}Z_{94})\Big\}  \nonumber \\
&& \Big\{ \Big[( s^2_h  Y_{62} + c_h  s_h  X_{62,63}Z_{64})
( s^2_h  Y_{79}Z_{78} + c_h  s_h  X_{79,80}Z_{78,81} + c_h  s_h  X_{79,91}Z_{78,98}  -  c^2_h  X_{80,91}Y_{79}Z_{78,81,98})\Big] \nonumber\\
&&-\ \Big[( c^2_h  X_{63,72}Y_{62}Z_{64} - c_h  s_h  X_{62,72})( s^2_h  Y_{79}Z_{78,81} + c_h  s_h  X_{79,80}Z_{78} + c_h  s_h  X_{79,91}Z_{78,81,98} 
\nonumber\\
&&-\ c^2_h  X_{80,91}Y_{79}Z_{78,98})\Big] \Big\}  R_X \, U_{4}[\theta_h] |0\rangle.
\end{eqnarray}

We can see that after the simplification, the expectation values of $W_{10}$, $W_{17}$ and $\tilde{W}_{17}$ can be rigorously calculated by exactly contracting the tensor network states with bond dimension $D=8, 8$ and $16$, respectively.
 The computational cost of exactly contracting these tensor network states is proportional to $\mathcal{O}(D^{10})$, and the memory cost is proportional to $\mathcal{O}(D^{8})$. We can save the memory space by the slicing trick~\cite{3DIsing,Pan}. It is worth noting that the expectation values of these stabilizers after depth reduction include many summation terms. The computational cost is very expensive if we separately calculate the expectation value of each term. Instead, we can utilize the summation trick of operations commonly used in DMRG~\cite{White} and 2D tensor network algorithms~\cite{Corboz2016,Laurens2016} to systematically absorb all relevant contributions into the left and right environment tensors, which greatly reduces the number of summation. These tricks allow us to obtain the exact expectation values of the $W_{10}, W_{17}$ and $\tilde{W}_{17}$ in less than 30 seconds, 30 seconds, and 5 hours per data point on a single Intel Xeon Gold 6326 CPU, respectively. In particular, we for the first time obtain the exact expectation value of the modified Weight-17 stabilizer as shown in Fig.~\ref{fig:shallow}, which can be used to benchmark the accuracy of other approximate methods. For more details, we refer to our source code in ~\cite{github}.

Similarly, the local magnetization $Z_{62}$ on a depth-T circuit can be reduced as follows
\begin{eqnarray}
\langle Z_{62} \rangle_{T} &=& \langle 0 | U^{\dagger}_{T}[\theta_h] (Z_{62}) U_{T}[\theta_h] |0\rangle \\
&=&  \langle 0 | U^{\dagger}_{T-1}[\theta_h] (c_h  Z_{62} +  s_h  Y_{62}) U_{T-1}[\theta_h] |0\rangle  \nonumber \\
&=& \langle 0 | U^{\dagger}_{T-2}[\theta_h] R^{\dagger}_X(c_h  Z_{62} +  s_h  X_{62}Z_{61,63,72}) R_X U_{T-2}[\theta_h] |0\rangle \nonumber \\
&=& \langle 0 | U^{\dagger}_{T-3}[\theta_h] R^{\dagger}_X \Big\{  c^2_h  Z_{62} + c_h  s_h  X_{62}Z_{61,63,72} -  c^3_h   s_h  Y_{62}  -  s^4_h  X_{61,62,63,72}Z_{60,64,81} \nonumber\\
&& +\ c_h   s^3_h  X_{63,72}Y_{62}Z_{64,81} + c_h   s^3_h  X_{61,72}Y_{62}Z_{60,81} + c_h   s^3_h  X_{61,63}Y_{62}Z_{60,64} +  c^2_h   s^2_h  X_{62,72}Z_{81} 
\nonumber\\
&&  + \ c^2_h   s^2_h  X_{62,63}Z_{64} +  c^2_h   s^2_h  X_{61,62}Z_{60} \Big\} R_X U_{T-3}[\theta_h] |0\rangle \nonumber\\
&=&  \langle 0 | U^{\dagger}_{T-4}[\theta_h] R^{\dagger}_X \Big\{ \big[  c^3_h  (1+ s^2_h ) Z_{62}  -  c^4_h  s_h  Y_{62} -  c^3_h  s^4_h  X_{62}Y_{61,63,72} \nonumber\\
&&  +\  c^2_h  s^3_h  X_{62}Z_{61,63,72}  - c_h  s^4_h  X_{61,62,63,72}Z_{60,64,81} -  s^7_h  X_{60,61,62,63,72}Y_{64,81}Z_{53,54,65} \big] \nonumber\\
&& -\ c_h  s^6_h  \big[ Y_{60,61,63,64}Z_{53,54,59,62,65} + Y_{60,61,72,81}Z_{53,59,62,80,82} + Y_{63,64,72,81}Z_{54,62,65,80,82} \nonumber\\
&& +\ X_{61,62,63}Y_{60,64,72}Z_{53,54,59,65}+X_{61,62,72}Y_{60,63,81}Z_{53,59,80,82}+X_{62,63,72}Y_{61,64,81}Z_{54,65,80,82} \big] \nonumber\\
&& +\  c^2_h   s^3_h  \big [ X_{61,63}Y_{62}Z_{60,64} + X_{63,72}Y_{62}Z_{64,81} + X_{61,72}Y_{62}Z_{60,81} \nonumber\\
&& +\ X_{61,62}Y_{60}Z_{53,59,63,72} + X_{62,63}Y_{64}Z_{54,61,65,72} + X_{62,72}Y_{81}Z_{61,63,80,82} \big] \nonumber\\
&& +\  c^2_h   s^5_h  \big [ (X_{61,62}Y_{60,63,72}Z_{53,59} + X_{62,63}Y_{61,64,72}Z_{54,65} + X_{62,72}Y_{61,63,81}Z_{80,82}) \nonumber\\
&& -\ (X_{63}+X_{72})Y_{60,61}Z_{53,59,62} - (X_{61}+X_{72})Y_{63,64}Z_{54,62,65} - (X_{61}+X_{63})Y_{72,81}Z_{62,80,82} \nonumber\\
&&-\ (X_{61,62,63}Y_{60,64}Z_{53,54,59,65,72} + X_{61,62,72}Y_{60,81}Z_{53,59,63,80,82} + X_{62,63,72}Y_{64,81}Z_{54,61,65,80,82}) \big ] \nonumber\\
&& +\  c^3_h   s^2_h  \big[ X_{61,62}Z_{60} + X_{62,63}Z_{64} + X_{62,72}Z_{81}
 - X_{62}Y_{72}Z_{61,63} - X_{62}Y_{63}Z_{61,72} - X_{62}Y_{61}Z_{63,72} \big] \nonumber \\
&& -\  c^3_h   s^4_h  \big[ X_{61,63}Z_{62} + X_{61,72}Z_{62}  + X_{63,72}Z_{62}
-X_{62,63}Y_{64}Z_{54,65}(Y_{61}Z_{72}+Y_{72}Z_{61}) \nonumber\\
&& -\ X_{61,62}Y_{60}Z_{53,59}(Y_{63}Z_{72}+Y_{72}Z_{63}) - X_{62,72}Y_{81}Z_{80,82}(Y_{61}Z_{63} + Y_{63}Z_{61})  \big] \nonumber \\
&& -\ c^4_h   s^3_h  \big[ X_{62}Y_{63,72}Z_{61} + X_{62}Y_{61,63}Z_{72} + X_{62}Y_{61,72}Z_{63}\big] \Big\} R_X U_{T-4}[\theta_h] |0\rangle
\end{eqnarray}
According to the reduced formula, the expectation values of the local magnetization at site 62 on the circuits with depth $n=3$ and $4$ have the following exact formulas
\begin{eqnarray}
 \langle Z_{62} \rangle_{3} &=&   c^3_h  \big( 1 +  s^2_h  \big), \\
 \langle Z_{62} \rangle_{4} &=&   c^4_h  \big( 1 + 2 s^2_h  - 3  c^2_h  s^{10}_h\big).
\label{Eq:Z62}
\end{eqnarray}

It is worth noting that if we exactly calculate $\langle Z_{62} \rangle_{4}$ on a depth-4 circuit by the time-evolution of the quantum state in the Schrodinger picture, we have to contract a tensor network with bond dimension $\chi = 64$. However, the exact formula Eq.~(\ref{Eq:Z62}) of $\langle Z_{62} \rangle_{4}$ is composed of $47$ Pauli-operator strings, thus the bond dimension is at most $47$ since each Pauli-operator string is corresponding to a product state. In practice, we only need to keep $\chi=5$ bond dimension to achieve machine precision for $\langle Z_{62} \rangle_{4}$. This reveals that our method can automatically find the Clifford low-rank structures.

In addition, at the Clifford point $\theta_h = \pi/2$, we have the properties of $R_X(\frac{\pi}{2})R_{ZZ}=R_{YY}R_X(\frac{\pi}{2})$ and $R_X(\frac{\pi}{2})R_{YY}=R_{ZZ}R_X(\frac{\pi}{2})$.  
Using these identities, we can express the original expectation value of $\langle Z_{62} \rangle$ as
\begin{eqnarray}
\langle Z_{62} \rangle_{20} &=& \langle 0| U^{\dagger}_{20}(\theta_h) Z_{62} U_{20}(\theta_h) | 0 \rangle \\ &=& \langle 0| U^{\dagger}_{19}(\theta_h) R^{\dagger}_{X} Z_{62} R_{X} U_{19}(\theta_h) | 0 \rangle  \\
&=& \langle 0| (\tilde{R}^{\dagger}_{X}R^{\dagger}_{YY}\tilde{R}^{\dagger}_{X})[R^{\dagger}_{ZZ}\tilde{R}^{\dagger}_{X}R^{\dagger}_{YY}\tilde{R}^{\dagger}_{X}]^{9} Z_{62} [\tilde{R}_{X}R_{YY}\tilde{R}_{X}R_{ZZ}]^{9}(\tilde{R}_{X}R_{YY}\tilde{R}_{X}) | 0 \rangle, 
\label{Eq:Rotation}
\end{eqnarray}
where $R_{X} = R_{X}(\theta_h) = \prod_{i}\exp \left(-{\rm i} \frac{\theta_{h}}{2}X_{i}\right)=R_{X}(\theta_h-\frac{\pi}{2})R_{X}(\frac{\pi}{2})=\tilde{R}_{X}R_{X}(\frac{\pi}{2})$, and $\tilde{R}_{X}=R_{X}(\theta_h-\frac{\pi}{2})$.

At the Clifford point $\theta_h =\pi/2$, the Heisenberg evolution operator in Eq.~(\ref{Eq:Rotation}) always has only a single Pauli-operator string, thus the bond dimension of PEPO is always 1. In contrast, if we perform the time evolution of the quantum state starting with the product state $|0\rangle$, the quantum state will become a maximally entangled state due to $R_{YY}|00\rangle = \frac{1}{\sqrt{2}}(|00\rangle - i|11\rangle)$, which makes the entanglement of the quantum state quickly grow. This is the reason why the MPS, isoTNS, and BP-TNS methods are difficult to compute the expectation values in the near-Clifford regime effectively.

\end{document}